\documentclass[12pt,preprint]{aastex}

\newcommand{\et}{et al.}
\newcommand{\kms}{km s$^{-1}$}
\newcommand{\ha}{H$\alpha$}

\newcommand{\sigtot}{$\sigma_{tot}$}
\newcommand{\sigz}{$\sigma_z$}
\newcommand{\sigr}{$\sigma_R$}
\newcommand{\sigphi}{$\sigma_\phi$}
\newcommand{\sigobs}{$\sigma_{obs}$}
\newcommand{\vhelio}{$V_{\rm helio}$}

\begin{document}

\title{The stellar velocity dispersion in the inner 1.3 disk scale-lengths 
of the irregular galaxy NGC 4449}

\author{Deidre A. Hunter\footnotemark[1]}
\affil{Lowell Observatory, 1400 West Mars Hill Road, Flagstaff, Arizona 
86001 USA}
\email{dah@lowell.edu}

\author{Vera C.\ Rubin\footnotemark[1]}
\affil{Carnegie Institution of Washington, 5241 Broad Branch Road, NW,
Washington, D.\ C.\ 20015 USA}
\email{rubin@dtm.ciw.edu}

\author{Rob A.\ Swaters}
\affil{
Department of Astronomy,
University of Maryland,
College Park, Maryland 20742-2421 USA}
\email{swaters@astro.umd.edu}

\author{Linda S.\ Sparke}
\affil{Washburn Observatory, 475 North Charter Street,
Madison, Wisconsin 53706-1582 USA}
\email{sparke@astro.wisc.edu}

\and

\author{Stephen E.\ Levine}
\affil{US Naval Observatory, Flagstaff Station, 10391 West Naval 
Observatory Road,
Flagstaff, Arizona 86001 USA}
\email{sel@nofs.navy.mil}

\footnotetext[1]{\rm Visiting Astronomer, Kitt Peak
National Observatory, National Optical Astronomy Observatory, which
is operated by the Association of Universities for Research in Astronomy, Inc.
(AURA) under cooperative agreement with the National Science Foundation.}

\begin{abstract}
We present measurements of the stellar velocity dispersion in 
the inner 1\arcmin\ radius 
(1.3 disk scale-lengths) 
of the irregular galaxy NGC 4449 
determined from long-slit absorption-line spectra. 
The average observed dispersion 
is 29$\pm$2 \kms, the same as predicted from NGC 4449's luminosity.
No significant rotation in the stars is detected. 
If we assume a maximum rotation
speed of the stars from the model determined from the gas kinematics
of Hunter et al. (2002),
the ratio $V_{max}$/\sigz\ measured globally is 3.
This ratio is comparable to values measured in spiral galaxies, 
and implies that the stellar disk in NGC 4449 is kinematically relatively cold.
The intrinsic
minor-to-major axis ratio $(b/a)_0$ is predicted to be
in the range 0.3--0.6, similar to 
values derived
from the distribution of observed $b/a$ of Im galaxies.
However, $V$/\sigz\ measured locally is 0.5---1.1,
and so the circular velocity of NGC 4449 is comparable or less than 
the velocity of the stars within the central 1.3 disk scale-lengths 
of the galaxy.
\end{abstract}


\section{Introduction} \label{sec-intro}

The intrinsic shape of Magellanic-type irregular (Im) galaxies
is controversial.
Studies of the distributions of projected optical minor-to-major axis ratios
$b/a$ have been used to infer the intrinsic shape
under the assumptions that there is one intrinsic shape and
that galaxies are oriented at random on the sky.
Hodge \& Hitchcock (1966) and van den Bergh (1988) concluded that
the intrinsic ratio $(b/a)_0$ of Im galaxies is
0.3--0.4 rather than the 0.2 value adopted for spirals.
\citet{sdk92}, on the other hand,
argue that the low ratio of
rotation velocity to velocity dispersion often seen in the gas
of irregulars must imply a thick disk with $(b/a)_0$
as high as 0.6. 
Yet others have interpreted the optical minor-to-major axis ratio
distribution to mean that Im galaxies
are triaxial in shape \citep{bp95},
only a little less spherical than 
dwarf ellipticals \citep{sung98}.

Fortunately, the stellar kinematics can be used as an indicator 
of the intrinsic shape of a galaxy,
and so we began a program of
observations of the stellar kinematics of Im galaxies.
In 2000 January we used the Kitt Peak National Observatory (KPNO) 
4 m Telescope with the RC spectrograph to
measure the stellar rotational velocities in
NGC 1156 and NGC 4449 \citep{rot02}.
In NGC 4449 we measured no organized rotation in the stars.
By contrast, clear organized rotation is
seen in the ionized and neutral gas.

NGC 4449 is not a pristine galaxy.
It has interacted with another galaxy 
to produce large streamers of HI that encircle the galaxy
(Hunter \et\ 1998). In addition,
the gas in the inner 2\arcmin\ of the galaxy is 
counter-rotating with
respect to the gas at larger radii, also the signature of
an interaction.
We explained the gas and stellar rotations with a model in which
the stars are in a disk seen nearly face-on, while the
gas lies
in a tilted disk with precession-induced twisting of the
line of nodes. 

This model is the simplest explanation that encompasses all the 
observations of NGC 4449.
However, it is also possibile that 
more of the kinetic energy of the stars
is in random motions rather than ordered rotation compared to the gas.
If this were the case, it could have profound implications for the 
shape of the stellar disk.

The ratio $V$/$\sigma$ is an indicator of how 
kinematically hot
a system is, and, therefore, an indicator of its structure.
Here $V$ is the speed of rotation and 
$\sigma$ is the velocity dispersion of the stars.
Elliptical galaxies and dwarf ellipticals (dE), 
which are triaxial systems, have $V/\sigma<1$ 
(Figure 4-6 in Binney \& Tremaine 1987; Bender, Saglia, \& Gerhard 1994;
Pedraz \et\ 2002; Pinkney \et\ 2003)
while spirals, which are cold thin disks, have $V/\sigma>1$, 
usually 2--5
\citep{bottema93,vega01}.

The stellar velocity dispersion has been measured in only two Im/Sm 
galaxies until now:
NGC 2552 (Sm; M$_B=-17.5$; Swaters 1999)
and the LMC (Im; M$_B=-18.1$; van der Marel \et\ 2002).
In both systems the vertical velocity dispersion of the
stars \sigz\ is
$\sim$20 \kms.
The relationship between integrated galactic magnitude M$_B$ and 
central velocity dispersion \sigz\ or \sigr\ for spirals and NGC 2552
(Bottema 1993, Swaters 1999)
predicts that the stars in galaxies with M$_B=-18.2$,
such as NGC 4449,
should have a velocity dispersion
of 29 \kms\ if most of the kinetic energy of the stars
is in ordered rotation.

Unfortunately, the spectral resolution of our earlier observations
was not adequate to resolve
the absorption profiles and allow a measurement of
$\sigma$. Therefore, we undertook new
observations of NGC 4449 with the 
Echelle spectrograph on the KPNO
4 m Telescope,
which have a resolution that is 2.3 times higher.
We present those observations here.
We adopt a distance of 3.9 Mpc to NGC 4449.

%Since NGC 4449 has interacted strongly enough to significantly disrupt
%the gas disk, one can consider whether NGC 4449 might be
%undergoing a transformation from an Im galaxy to a 
%puffy dE. We will see below that the stellar velocity
%dispersion of NGC 4449 is in the range of what is observed in 
%dEs (14--39 \kms; Peterson \& Caldwell 1993).
%Furthermore, similarity of velocity gradients between some
%dEs and dIms of the same luminosity
%has led to the suggestion that dEs might be
%stripped dIms (van Zee, Skillman, \& Haynes 2004).
%Thus, it is particularly relevant to investigate the kinematics
%of NGC 4449's stars and what this might tell us about its
%structure.

\section{Observations and Data Reduction} \label{sec-obs}

We observed NGC 4449 with the Echelle spectrograph on the KPNO
4 m Telescope during three nights in 2003 May. We used the
UV camera, a Tektronics $2048\times2048$ CCD, and the 316--63\arcdeg\ echelle
grating. The cross-disperser
was replaced with a silver flat in order to obtain a
useful slit length of 3\arcmin, and the slit width was 2.5\arcsec.
%The FWHM of the comparison lines was 22 \kms.
We used a narrow-band
post-slit filter (KPNO \#1433, central wavelength of 5204 \AA\ and
FWHM of 276 \AA) and
GG495 as a pre-slit filter. The
final spectrum covered 5050 \AA\ to 5287 \AA\ at 0.135 \AA\ per pixel.
The spectra included, therefore, the 
MgIb lines at 5167 \AA, 5173 \AA, and 5184 \AA, which were the strongest
absorption lines in the spectra.
We binned by two in the spatial direction for a pixel scale of 1.22\arcsec.

NGC 4449 was observed at three position angles (PA): 46\arcdeg, the
morphological major axis; 316\arcdeg, the minor axis;
and 271\arcdeg, plus 45\arcdeg\ from the major axis.
These are 3 of the 4 PAs (modulo 180\arcdeg) that we 
used in our observations with the RC spectrograph. 
The slit was placed across the center of NGC 4449, 
determined from V-band images,
by offsetting from a nearby
star. However, 
we then nudged the position of the slit 
4.3\arcsec\ south in PA 316\arcdeg\ and 0.2\arcsec\ east and 1.8\arcsec\
north in PA 271\arcdeg\ to move the
slit off the bright super star cluster near the galaxy center.
Slit positions are drawn on
V-band and \ha\ images of the galaxy in Figure \ref{fig-vha}.
%There was some cirrus during observations of PA 271\arcdeg.
We took 5--6$\times$1800 s spectra at each PA.
Because NGC 4449 is bigger than the slit, we offset
to a separate position to measure sky,
%The offset position was about 8\arcmin\ from the
%center of NGC 4449 and was chosen by looking for a clean spot on
%an extract of the Digitized Sky Survey. 
%Because the moon was near
%full for these observations, sky was an important component of the spectra
%and had to be determined very well. Therefore, 
and we sandwiched 
each galaxy observation
between sky observations of equal integration time.

To remove pixel-to-pixel variations, we 
observed the white spot mounted inside the dome. 
The overscan portion of the CCD
image was used to remove the electronic pedestal.
Observations of Th-Ar comparison lamps 
were used to set
the wavelength scale and map spatial distortions along the slit.
A star observed at intervals along the slit was used to remove 
S-distortions. Finally, we observed 6--8 radial velocity
standards of type F, G, and K each night from the \citet{nauti03},
and these were used as cross-correlation
templates. 
Twilight sky 
provided a template of a G2V star. 

The two-dimensional spectra were repixelized
in order to impose a uniform wavelength
scale and to correct 
for curvature along lines of constant wavelength
and along lines of constant spatial position.
We transformed to a logarithmic wavelength scale,
%We examined individual spectra taken at the same PA
%to make sure that the galaxy features
%were aligned along the slit.
subtracted sky,
and combined the multiple images, rejecting
cosmic rays. 
%For the stars, sky was determined on either side of 
%the star spectrum.

\section{Cross-correlation Measurements} \label{sec-meas}

Two of us undertook
independent measurements of the heliocentric radial velocities and
velocity dispersions in the spectra. 
In both approaches, one-dimensional galaxy and stellar spectra were 
extracted along the slit, and the
continuum was fit and subtracted
from each spectrum.
The one-dimensional spectra were then cross-correlated against each of
the stellar spectra observed on the same night, first the stars
against themselves to look for systematic problems and then the galaxy
spectra with the stars. 
We used the spectrum from 5134--5198 \AA\ and 5205--5234 \AA, excluding
the region 5198--5205 \AA\ because of the presence of
galactic emission lines.
Low frequencies resulting from residual continuum were removed with a
step-function Fourier filter, and the
peak of the cross-correlation profile was fit with
a Gaussian.

The two approaches varied in
the details of continuum fitting and Fourier filtering.
In addition, in the first approach, D.A.H. extracted one-dimensional spectra 
summed every 24.4\arcsec.
In the second approach, R.A.S. used one-dimensional spectra 
that had been boxcar-smoothed by 12.2\arcsec.
Most of the measurements of the two approaches agree
within the uncertainties. A few radial positions that do not agree within the
error bars are undoubtedly due to the differences in sampling 
and the other details
of the approaches, and perhaps better represent the overall uncertainties.
We present the results of both measurements below: Approach 1 is plotted
in blue and Approach 2 is plotted in red in the plots that follow.

The results of the cross-correlation---heliocentric 
velocity \vhelio\ and
FWHM of the profiles---obtained with all of the stars were averaged,
regardless of the stellar type of the template.
The uncertainty of each \vhelio\ and FWHM was taken as the
combination of the dispersion around the mean and the uncertainty
in the fit, summed in quadrature, but the uncertainty
in the fit dominates. 
%the cross-correlation routine used in 
%Image Reduction and Analysis Facility (IRAF) does not
%provide a fit uncertainty. Therefore, the cross-correlation profiles
%from the first approach described above
%were written out and fit in Astronomical Image Processing Software
%(AIPS) to determine the uncertainty
%in the FWHM. The final uncertainty in the FWHM was then taken
%as the quadratic sum of the fit uncertainty and the dispersion
%around the mean.

We measured an average central velocity 
of 205$\pm$1 \kms, where
the uncertainty represents the night-to-night variation.
The zero point of each night has been adjusted to the 
three-night average of the 
central velocity.
In \citet{rot02} the central velocity measured from
the spectra was 214 \kms. To compare the current measurements
with those made in 2002, we subtracted 9 \kms\ from each
2002 heliocentric velocity to give them the
same central velocity as the current measurements.

To determine the observed velocity dispersion of the stars
\sigobs, the FWHM measured from the galaxy 
profiles were corrected for the resolution given by cross-correlation
of the template stars against each other, 
assuming that the FWHM add in quadrature.
The average FWHM of the template stars was 47$\pm$1 \kms\
for the measurements in Approach 1 and 32 \kms\ in Approach 2.
%The value of \sigobs\ is FWHM/2.35.

\section{Cross-correlation Results} \label{sec-results}

\subsection{Stellar rotation} 

Figure \ref{fig-rot} shows the resulting heliocentric velocity
\vhelio\ as a function of position
from the center of the galaxy. 
Positive numbers denote position along the slit in the direction
of the given PA.
Included are the
data from \citet{rot02}.
Our new data cover approximately
the inner 1\arcmin\ radius. This is 1.1 kpc at a distance of 3.9 Mpc
or 1.3 optical disk scale-lengths, where the scale-length of 0.84 kpc is
an exponential fit to 
V-band surface photometry measured by Hunter \et\ (1999).

\citet{rot02} reported no detectable rotation from the stars
at any PA with an upper limit of ($3/\sin i$) \kms\
kpc$^{-1}$. 
Our current measurements largely confirm this lack of observed rotation. 
However, at PA 271\arcdeg\ 
the measurements plotted in red in Figure \ref{fig-rot} do suggest
a velocity gradient in the stars,
and this gradient is intriguingly close to that measured for the
ionized gas. A least-squares fit to the red points gives
a velocity gradient of $9\pm2$ \kms\ kpc$^{-1}$.
The measurements plotted in blue, which average over
twice the spatial scale, 
lie within the uncertainties near the measurements plotted in red,
but do not themselves show a velocity gradient.
A fit to both sets of measurements yields a gradient that is
half that seen in only the measurements plotted in red.
Also, we do not see any hint
of rotation at the other two PAs. By contrast, the amplitude of
the velocity gradient for the ionized gas 
in our 2002 results was largest at a PA of 46\arcdeg.

However, the uncertainties in the stellar \vhelio\ at the three PAs
are large and encompass several
possibilities including no rotation of the stars
or weak rotation 
with almost any kinematic major axis provided the 
inclination is not too large. For example, with the
inclination of 68\arcdeg\ and a kinematic major axis
of 80\arcdeg\ from our model for the kinematics of the inner HI gas,
the gradient seen at PA 271\arcdeg\ transformed to the other
PAs would lie within the uncertainties of the \vhelio.

Below we will discuss two limiting cases. 
In the first case the lack of
rotation in the stars is taken as due to viewing the
stellar disk face-on (Hunter \et\ 2002).
This would imply that the stars and gas
lie in different planes.
In the second case, we will assume that the  
stars are rotating 
at an inclination angle (68\arcdeg) and a kinematical major axis
(80\arcdeg) that are the same as those of the HI gas.

There is some indirect evidence that the stars are not in the 
same plane as the neutral gas.
In V-band, NGC 4449 has a boxy shape that becomes round
in the outer parts, and there is a twisting of
the isophotes from the central rectangle to the outer galaxy
(see Figure 1).
Hunter \et\ (1999)
interpret these characteristics as a bar structure that
has a length of
3.9 disk scale-lengths.
The shape of this bar is symmetrical in V, and to 
deproject this structure by a large angle ($>$30\arcdeg)
would produce a sharp-angled structure not seen in other IBm galaxies
(Hunter \& Elmegreen 2005)
unless the line of nodes runs along the bar's major or minor axis.
Our model for the inner gas disk has the line of nodes
at PA 80\arcdeg, almost halfway between the long and short axes of the bar.
Thus, the stellar disk needs to be closer to face-on than the gas
or tilted about a different axis.
Furthermore, the bar in NGC 4449 has an apparent $b/a$ of
0.6, a value that is typical of bars in IBm galaxies (Hunter \& Elmegreen 2005).
Deprojecting the bar by a large angle would produce an axis
ratio that is not seen in IBm galaxies.
Finally, if the bar in NGC 4449 is rotating, then the rotation speed
along the bar's minor axis (PA of 316\arcdeg) should be higher than the
circular speed. To hide rotation at PA 316\arcdeg,
the stellar disk would need to be tilted about PA 46\arcdeg.
On the other hand, we do detect stellar absorption to the greatest
extent along PA 46\arcdeg\ and to the least extent along PA 316\arcdeg.
This is similar to the ionized gas which was best fit with
a kinematic major axis
at PA 46\arcdeg.
Thus, the two cases we are considering---a face-on disk and 
a disk inclined
at 68\arcdeg\ with a PA of 80\arcdeg---most likely bracket
the true orientation of the stellar disk.

\subsection{Stellar velocity dispersion}

Figures \ref{fig-fwhm} and \ref{fig-fwhmcomb}
show the observed velocity dispersions \sigobs.
No trend with radius is clear
except in PA 46\arcdeg\ where
the lowest \sigobs\ fall on the southwest side of the galaxy.
Furthermore, differences in average \sigobs\ between PAs 
(1--4 \kms) are smaller than or comparable to
the uncertainties (3--4 \kms).
Therefore, we have averaged all of the measurements of \sigobs\
at all PAs together, weighted by uncertainty, and averaged
the results of the two approaches to the measurements.
The final average \sigobs\
is 29$\pm$2 \kms.

\subsubsection{Case 1: Face-on Stellar Disk}

In the case where we are looking at the stellar disk nearly face-on,
\sigobs\ is simply the vertical component \sigz\ of \sigtot,
where the total velocity dispersion \sigtot\ is the
quadratic sum of the vertical component \sigz, the radial component
\sigr, and the azimuthal component \sigphi.
From the relationship between M$_B$ and
\sigz\ (Bottema 1993, Swaters 1999), 
we predict a \sigz\ of 29 \kms\ for NGC 4449. Thus, 
our observed \sigz\ is the same as predicted by the galaxy's
luminosity.

In normal, well-mixed, axisymmetric disk systems in which the 
vertical scale height
does not change with radius, the velocity dispersion drops with
radius as the square-root of the surface density.
In Figure \ref{fig-fwhmcomb} one can see that in NGC 4449
at R$\geq 56$\arcsec\ at PAs 46\arcdeg\ and 271\arcdeg\ the
\sigobs\ are in fact lower than those in the central region.
The average of \sigobs\ at radii less than 20\arcsec\ in all PAs is 
31$\pm$6 \kms, where the uncertainy of 6 \kms\ represents the
range in values observed in the central region. If NGC 4449
were a well-behaved disk, we would then predict a \sigobs\
of 14$\pm$3 \kms\ at a radius of 69\arcsec. The average
\sigobs\ at radii 56\arcsec\ to 81\arcsec\ is 19$\pm$4 \kms,
where again the uncertainty represents the range of values
measured there. Although the uncertainties are large, 
we see that, within the uncertainties,
the drop in \sigobs\ in the outer parts of the region we
have observed is consistent with a normal face-on stellar disk.

\subsubsection{Case 2: Inclined Stellar Disk}

If the stellar disk is inclined, our
\sigobs\ measured along some position angle 
is a combination of the three components of the
velocity dispersion: 
$$\sigma_{obs}^2=(\sigma_R^2\sin^2 \eta + \sigma_\phi^2\cos^2 \eta)\sin^2 i +
\sigma_z^2\cos^2 i,$$
where $\eta$ is the angle between the observed PA and 
the major axis and $i$ is 
the inclination angle of the disk (e.g., Gerssen \et\ 1997).

In the Milky Way and other spirals,
the ratios of these three components are approximately constant
with radius within a galaxy and 
similar between galaxies. 
Gerssen, Kuijken, 
\& Merrifield (1997) measured \sigz/\sigr$\sim0.7\pm0.2$ in
the Sb galaxy NGC 488. From the data of Delhaye (1965) as
presented by Binney \& Merrifield (1998),
giant stars in the solar neighborhood have \sigz/\sigr$\sim0.73\pm0.09$.
Other stellar components have similar ratios.
For \sigphi\ and \sigr,
the nearby red giants give a ratio of \sigphi/\sigr$\sim0.65\pm0.08$.
Similarly, the epicyclic approximation and the
observational determination of the Oort constants lead
to the relation 
$\sigma_\phi^2/\sigma_R^2 = 0.5 (1 + {{d \ln v} \over {d \ln R}})$
(Binney \& Tremaine 1987, Binney \& Merrifield 1998), where
$v$ is the rotation speed at radius $R$,
or \sigphi$\sim0.7$\sigr\ where the galactic rotation curve is flat.
These together suggest the approximate relationships
\sigz$\sim$\sigphi$\sim$0.7\sigr.

How well these relations apply to NGC 4449 is not clear since
this galaxy is dominated by a very large bar potential as identified
in the optical by Hunter \et\ (1999) and discussed in \S4.1 above.
This could result in non-circular motions, although
such are not easily identified in the gas due to the complexity 
of the counter-rotation in the inner parts.
However, to the extent that these relations do apply to NGC 4449,
then along the major axis the dependence on the inclination
drops out and \sigobs$\sim$\sigz, but
at any other PA, a dependence on the inclination remains.
For an inclination of 68\arcdeg\ and kinematic major axis
of 80\arcdeg,
\sigobs\ is 1.02, 1.13, and 1.27 times \sigz\ at the
three PAs of the observations. The minimum value that we observe
should occur at a PA of
271\arcdeg\ and the maximum at PA 316\arcdeg.
This is opposite to what we see
(31$\pm$3 \kms\ at PA 271\arcdeg\
and 27$\pm$4 \kms\ at PA 316\arcdeg), but the uncertainties are large.
Thus, \sigz\ is most likely between 23 \kms\ and 28 \kms, for
a \sigobs\ of 29 \kms.
We adopt the average, and thus,
\sigz\ is 25 \kms\ for this case.

\section{Discussion}

Because we wish to define the structure of NGC 4449 from its stellar
kinematic properties, we must evaluate
the ratio of the rotation
speed to the velocity dispersion $V$/$\sigma$.
Since we wish to compare 
$V$/$\sigma$ measured in NGC 4449 to values measured in other
galaxies, it is important that we measure $V$ and $\sigma$
in the same way.
For elliptical galaxies, Bender \et\ (1994) used \sigobs\ averaged
within a radius of 0.5R$_e$, where R$_e$ is the effective radius,
and Pinkney \et\ (2003) used \sigobs\ measured in the center of
early type galaxies.
Bottema's (1993) observations of a sample of spirals indicate
that \sigr\ at a radius of one scale-length is similar in value
to \sigz\ at the center of the galaxy. They used both of these
quantities.
\citet{vega01} give \sigobs\ at 0.25R$_e$ for spirals.
Thus, the quantity that makes the
most sense to use here is \sigz\ measured in the central region
of NGC 4449.
For $V$ in the ratio $V$/$\sigma$, all of the references
mentioned above
use the maximum rotation speed $V_{max}$ of the system. 
Therefore, we should use $V_{max}$/\sigz\ for 
NGC 4449. 

Unfortunately, we did not reliably measure rotation 
in the stars.
However, the gas does show clear rotation.
Even if the stars and gas lie in different planes, 
we would expect that they
see nearly the same gravitational potential.
That this is a reasonable assumption 
is shown in observations of polar ring galaxies where
the maximum rotation speed in the disk is the same as
that in the ring to within 25--35\% 
(Sackett \et\ 1994, Swaters \& Rubin 2003).
Therefore, it seems likely that the maximum rotation 
speed of the stars in NGC 4449 is no more than $V_{max}$
for the gas, which was 80 \kms\ in our model.

The uncertainty in the ratio $V_{max}$/\sigz\ is dominated by the uncertainty
in the stellar $V_{max}$, not in the effect of inclination of the disk
on the measurement of \sigz.
Thus, for both of our cases,
a $V_{max}$ of 80 \kms\ implies a
$V_{max}$/\sigz\ of 3. 
For spiral galaxies $V$/$\sigma$ is measured to be 2--5
while for ellipticals, this ratio is less than one.
Therefore, NGC 4449's value of $V$/$\sigma$
is like those measured in spiral galaxies and 
implies that NGC 4449 also has a kinematically relatively cold disk.
In addition, since \sigz\ drops radially in disks, it is unlikely
that the stars further out in the disk of NGC 4449 have a higher
velocity dispersion than those in the central region, and
Figure \ref{fig-fwhmcomb} seems to be consistent with this.

Of course, the rotation speed of the stars is not known,
and it could be {\it lower} than that of the gas if more
kinetic energy is going into random motions than into ordered rotation.
The fact that we measured a value for \sigz\ that was predicted for NGC 4449
by its M$_B$ suggests that the velocity dispersion of the stars
is not extraordinarily high.
Furthermore, the velocity dispersion of the gas within 
the optical part of NGC 4449 (20 \kms, Hunter \et\ 1999)
is comparable
to the stellar velocity dispersion, making it unlikely that there
is an appreciable asymmetric drift of the stars relative to
the mean rotation velocity of the gas.
For NGC 4449 to have a ratio $V$/$\sigma$ that is minimally comparable
to that of triaxial systems, the stellar $V_{max}$ would have
to be one-third that of the gas. 

For an isothermal stellar population with a constant velocity dispersion
in a dark halo, in the outer galaxy $h_z/R \sim \sigma_z/v_c$ 
where $v_c$ is the circular velocity and $h_z$ is the disk scale-height.
On the other hand, for a self-gravitating disk with constant velocity,
near the center of the disk $h_z/R \sim (\sigma_z/v_c)^2$
(Binney \& Merrifield 1998).
This suggests that the intrinsic minor-to-major axis ratio $(b/a)_0$
of the NGC 4449 stellar disk lies in the range 0.3--0.6. A value of 0.3--0.4
has been deduced for Im
galaxies by Hodge \& Hitchcock (1966) and van den Bergh (1988)
from studies of observed $b/a$ distributions, and a value of 0.6 
has been suggested by 
Staveley-Smith \et\ (1992).

There are two other Im/Sm galaxies that have measures of $V_{max}$/\sigz.
For the LMC, this ratio is 3 (van der Marel \et\ 2002; see
also Nikolaev \et\ 2004),
and in NGC 2552 it is 5 \citep{swaters99}.
Thus, these galaxies also have ratios that 
imply that they have
flattened kinematically cold disks.
However, we note that
the three galaxies that have now been observed are at the high
end of the range of M$_B$ of Im galaxies, in the region that overlaps
with late-type spirals.
Perhaps more typical, lower luminosity Im galaxies are shaped differently.

It is also of interest to examine the local value of the ratio $V$/\sigz.
For NGC 4449, our measurement of \sigobs\ applies 
to the inner 1\arcmin.
In our model of the HI kinematics, 
the circular velocity ramps
linearly from 0 to 80 \kms\ between the center and 2.5\arcmin.
Thus, the maximum rotation speed at a radius of 1\arcmin\
is about 32 \kms.
We experimented with tweaking the model to see if the
circular velocity could remain high closer into the center,
and found that such was not consistent with the HI observations.
So, for the case of a stellar disk seen nearly face-on, 
($V$/\sigz)$_{R=1\arcmin}$ is 1.1.
If the stars are rotating in a disk seen 
at an inclination of 68\arcdeg, the observed gradient discussed
in Section \ref{sec-results} implies a rotation speed of
only 12 \kms\ at a radius of 1\arcmin. With a \sigz\ of
25 \kms, the ratio ($V$/\sigz)$_{R=1\arcmin}$
is 0.5.
In either case the mean rotation velocity is comparable to or less
than the
velocity dispersion of the stars in the central 1.3 disk
scale-lengths of NGC 4449.

\section{The future of NGC 4449}

In Hunter et al. (2002) we constructed a simple model for the
distribution of HI and ionized gas in NGC 4449, in which the gas is in orbits of
constant inclination to the plane of the sky, but with
precession-induced twisting of the line of nodes.  Dynamically, this
is what we would expect for a tilted gas disk precessing in a
gravitational potential that is roughly axisymmetric and flattened
with its midplane in the plane of the sky (as the stellar disk appears
to be).  What does this model imply for the current age and the future
development of the twisted gas disk?

Our model gas disk is inclined at 68\arcdeg\ to face-on, and
twists through 180\arcdeg\ about the line of sight.
To fit the HI measurements, we took the rotation curve to rise linearly 
to 80 \kms\ at 2.5\arcmin\ or 3 kpc from the center.  Thus, within 
$R = 2.5$\arcmin, the angular rate of rotation 
$\Omega_{rot} = [80 {\rm km/s}]/[3~{\rm kpc}] \approx 27~{\rm Gyr}^{-1}$ 
and the rotation period is 250~Myr.  Beyond that the rotation curve is 
flat, so the rotation rate $\Omega_{rot}$ is proportional to $1/R$.

An orbit tilted at angle $\theta$ to the equatorial plane of the potential
precesses at a rate $\Omega_{rot} \epsilon_{\Phi} \cos(\theta)$, where
$\epsilon_{\Phi}$ is the flattening of the equipotential contours.  From
Section 2.2.2 of Binney \& Tremaine (1987), we expect the equipotentials 
to be roughly a third as flattened as the density contours.  The total 
mass distribution is unlikely to be {\it more} flattened than the stellar  
distribution, which has an axis ratio $b/a=0.3 - 0.6$; thus 
$\epsilon_{\Phi} \approx 0.33 (1-b/a) = 0.1 - 0.2$.  Setting 
$\theta = 68$\arcdeg, this corresponds to a precession period 
of 3--7~Gyr within 2.5\arcmin, decreasing as $1/R$ at larger radii.

Our model fits the HI distribution within 12\arcmin.  The gas orbits
twist by 190\arcdeg\ between radii of 2\arcmin\ and 5\arcmin, and have
a constant orientation further out.
Our model is the simplest that fits the HI velocities, and more
likely the real twist is more gradual.
If the entire gas disk was
originally coplanar, it would take 2--3 Gyr to develop a twist of
190\arcdeg\ between gas orbiting at 2\arcmin\ and that at 10\arcmin.
If we take this as a very rough indication of the structure's age, it
is consistent with the dynamical model of Theis \& Kohle (2001) for
the outer gas disk.  They suggest (Section 4.2) that the HI disk may
have originated in an encounter with neighbor NGC~4736 roughly 4~Gyr
ago; a close passage by the dwarf DDO~125 0.5~Gyr ago perturbed the
gas disk but is unlikely to have been its origin.

In another 2--3~Gyr, the part of the disk within 2\arcmin\ will have
twisted by 360\arcdeg\ relative to that beyond 5\arcmin.  Beyond that
time, we might no longer recognize the entire structure as a
smoothly-twisted disk.  Instead, we would probably interpret it as a 
broad HI ring, with disconnected clouds of gas in variously inclined 
orbits nearer to the galaxy center.  

\acknowledgments

We are grateful to the people at KPNO who helped make these observations
successful, and especially to Daryl Willmarth who recommended a more
efficient
instrument setup than we had originally planned. We are also extremely
thankful to Phil Massey who
assisted in the observations.
L.S.S. is grateful for hospitality at the Max-Planck-Institute for 
Astrophysics in Garching.
Funding for this work was provided in part by
the National Science Foundation through grants AST-02-04922 to D.A.H.
and AST-00-98419 to L.S.S.

Facilities: \facility{Lowell Observatory}, \facility{KPNO}.

\clearpage

\begin{figure}
\epsscale{.80}
%\plotone{f1.eps}
\caption{
False-color representation of the logarithm of the
V-band and \protect\ha\ images of NGC 4449 (Hunter et al.\ 1999).
In both images lines are drawn along the three observed
position angles PA: 46\arcdeg, 316\arcdeg, and 271\arcdeg.
The extent of the lines indicates the region of the slit over
which measurements of the stellar velocity dispersion and heliocentric
velocity were made.
The circles mark the center of the galaxy.
\label{fig-vha}}
\end{figure}

\clearpage

\begin{figure}
\epsscale{0.5}
%\plotone{f2.eps}
\caption{Stellar line of sight heliocentric radial velocities
\protect\vhelio\ measured from absorption lines in each spectrum
of NGC 4449. 
The two approaches to measuring the spectra are
discussed in Section \protect\ref{sec-meas}.
The position from the center of the galaxy
is measured along the slit, and positive numbers refer to the direction
given by the position angle PA.
The solid horizontal line is the adopted central velocity
of 205 \protect\kms. The dashed
line is the fit to the rotation velocities of the ionized gas
at PA 46\arcdeg\ by \citet{rot02}.
The measurements of 
Hunter et al.\ (2002) are also included,
with their results at 136\arcdeg\ and 91\arcdeg\ 
flipped to match the PAs
used here. An offset of 9 \protect\kms\ has been
subtracted from the Hunter et al.\ measurements to account
for the difference in central velocity that was used.
The PAs corresponding to the major and minor axes of the bar are noted.
\label{fig-rot}}
\end{figure}

\clearpage

\begin{figure}
\epsscale{0.5}
%\plotone{f3.eps}
\caption{Observed stellar velocity dispersion \protect\sigobs\ of absorption lines
measured from the cross-correlation profiles
for each spectrum of NGC 4449. The solid horizontal line
denotes the uncertainty-weighted average of all of the measured dispersions.
The \protect\sigobs\ are determined from the cross-correlation FWHM and
corrected for the intrinsic
resolution as given by the template stars.
The two approaches to measuring the spectra 
discussed in Section \protect\ref{sec-meas} are plotted in different colors.
The PAs corresponding to the major and minor axes of the bar are noted.
\label{fig-fwhm}}
\end{figure}

\begin{figure}
\epsscale{1.0}
%\plotone{f4.eps}
\caption{Same as Figure \protect\ref{fig-fwhm}, but
measurements of the observed velocity dispersion \protect\sigobs\
along different position angles PA are plotted 
together as a function of distance from the center of the galaxy.
PA 46\arcdeg\ is the morphological
major axis of the bar and PA 316\arcdeg\ is the
bar minor axis.
The two approaches to measuring the spectra 
discussed in Section \protect\ref{sec-meas} are plotted in different colors.
\label{fig-fwhmcomb}}
\end{figure}

\end{document}